\documentclass[manuscript]{aastex631_arxiv} 

\usepackage{color}
\usepackage{graphicx}
\usepackage{bm}
\usepackage{epsf}
\usepackage{wrapfig}
\usepackage{multirow}
\usepackage{epstopdf}
\usepackage{tabularx}
\usepackage{natbib}
\usepackage{amssymb,amsmath,mathtools,amsthm}
\usepackage{mathrsfs}
\usepackage{nccmath}
\usepackage[thinlines]{easytable}
\usepackage{hhline}

\begin{document}

\title{Interaction between a Coronal Mass Ejection and Comet 67P/Churyumov-Gerasimenko}

\correspondingauthor{Zhenguang Huang}
\email{zghuang@umich.edu}

\author{Zhenguang Huang}
\affiliation{Climate and Space Sciences and Engineering, University of Michigan, Ann Arbor, MI 48109, USA}

\author{G\'abor T\'oth}
\affiliation{Climate and Space Sciences and Engineering, University of Michigan, Ann Arbor, MI 48109, USA}

\author{Tamas I. Gombosi}
\affiliation{Climate and Space Sciences and Engineering, University of Michigan, Ann Arbor, MI 48109, USA}

\author{Michael R. Combi}
\affiliation{Climate and Space Sciences and Engineering, University of Michigan, Ann Arbor, MI 48109, USA}

\author{Xianzhe Jia}
\affiliation{Climate and Space Sciences and Engineering, University of Michigan, Ann Arbor, MI 48109, USA}

\author{Yinsi Shou}
\affiliation{Climate and Space Sciences and Engineering, University of Michigan, Ann Arbor, MI 48109, USA}

\author{Valeriy Tenishev}
\affiliation{Climate and Space Sciences and Engineering, University of Michigan, Ann Arbor, MI 48109, USA}

\author{Kathrin Altwegg}
\affiliation{Space Research \& Planetary Sciences, Physics Institute, University of Bern, Bern, Switzerland}

\author{Martin Rubin}
\affiliation{Space Research \& Planetary Sciences, Physics Institute, University of Bern, Bern, Switzerland}

\begin{abstract}
The interaction between a Coronal Mass Ejection (CME) and a comet has been observed several
times by in-situ observations from the Rosetta Plasma Consortium (RPC),
which is designed to investigate the cometary magnetosphere of comet 67P/Churyumov-Gerasimenko (CG).
\cite{Goetz_2019} reported a magnetic field of up to 300\,nT measured in the inner coma, which is among the largest interplanetary magnetic fields observed in the solar system.
They suggested the large magnetic field observations in the inner coma
come from magnetic field pile-up regions, which are generated by the interaction between 
a CME and/or corotating interaction region and the cometary magnetosphere. However, the detailed interaction
between a CME and the cometary magnetosphere of comet CG in the
inner coma has not been 
investigated by numerical simulations
yet. In this manuscript, we will use a numerical model to simulate
the interaction between comet CG and a Halloween class CME 
and investigate its magnetospheric response to the CME. We find that
the plasma structures change significantly during the CME event, and the maximum value of the magnetic field strength is more than 500\,nT close to the nucleus. Virtual satellites at similar distances as Rosetta show that the magnetic field strength can be as large as 250\,nT, which is slightly less than what \cite{Goetz_2019} reported. 

\end{abstract}




\section{Introduction}
The Rosetta Plasma Consortium (RPC, \cite{Carr_2007}) onboard 
Rosetta \citep{Glassmeier_2007_Rosetta} was designed to
study the physical and chemical properties of comet 67P/Churyumov-Gerasimenko (CG), 
the evolution of the cometary ionosphere and magnetosphere,
as well as examining how the interaction region of the solar wind and the comet is developing. 
The investigation of the cometary plasma environment and the
locations of plasma boundaries for an active comet 
started with the Giotto flyby
of comet 1P/Halley. Different plasma boundaries have been identified, such 
as a bow shock, which forms upstream of the comet where
the solar wind slows down from supersonic to subsonic speeds \citep{Galeev_1985, Koenders_2013}. And if the comet is very
active with a high neutral gas production rate near perihelion, 
a diamagnetic cavity (inside which the magnetic field drops to
zero) is formed because the outward pointing ion-neutral drag force
is large enough to balance the inward-pointing magnetic force 
(the sum of the magnetic pressure gradient and magnetic curvature forces)
\citep{Neubauer_1986, Cravens_1986}. Recent studies by \cite{Goetz_2016, Goetz_2016_2} and \cite{Henri_2017} suggested that instabilities may extend the
size of the diamagnetic cavity to become larger than expected from numerical simulations or theoretical predictions.
Besides, an inner shock, where the cold supersonic cometary ion outflow (from the surface
of the nucleus) slows down to subsonic velocities (as it reaches near zero velocity near the contact surface
along the sun-comet line), is also proposed \citep{Gombosi_2015}. 

The Rosetta mission has dramatically improved our understanding 
of the cometary plasma environment
since its arrival at comet CG in 2014. One of the most important 
discoveries is that the diamagnetic cavity is somewhat different from 
our previous understanding of a global diamagnetic
cavity \citep{Goetz_2016, Goetz_2016_2}. 
Many small-scale structures along the diamagnetic 
surface are reported, suggesting that the diamagnetic cavity of comet CG differs from a traditional, global-scale diamagnetic 
cavity (like at comet 1P/Halley). The small-scale structures 
are believed to arise from plasma waves along the global diamagnetic cavity 
boundary \citep{Goetz_2016, Goetz_2016_2}. The electron environment near
the diamagnetic cavity has been extensively studied  \citep{Henri_2017,Odelstad_2018,Engelhardt_2018}. Readers can also
refer to the review article by \cite{Goetz_2022}.
\cite{Huang_2018} provided an alternative explanation
that the Hall current along the cavity boundary can introduce magnetic reconnection on the day-side and generate weak magnetic field regions outside the global diamagnetic cavity, similar to the RPC observations. 
Besides, these shallow magnetic fields were observed much farther from
the nucleus than the predicted diamagnetic cavity locations from
numerical simulations \citep{Rubin_2015, Koenders_2015,
Huang_2016}. \cite{Mandt_2016} noticed a `mystery' boundary 
separating an inner region and an outer region,
which is not predicted by previous studies \citep{Gombosi_2015,
Rubin_2015, Koenders_2015, Huang_2016}.  They suggested that
an ion-neutral collisionopause
boundary, inside of which collisions between ions and neutrals are
important and outside of which they are not, is responsible for forming the `mystery' boundary.
Recently, \cite{Gunell_2018} reported that an infant bow shock was observed several times a few months before and after perihelion, with signatures
of an increase in the magnetic field magnitude and fluctuations, an increase in the electron and proton temperatures 
at the shock and the diminution of the solar wind plasma downstream.

Space weather-like events in comet CG's magnetosphere have also been reported by
the Rosetta mission \citep{McKenna-Lawlor_2016, Edberg_2016JGR, Edberg_2016MNRAS, Noonan_2018, 
Goetz_2019}. \cite{McKenna-Lawlor_2016} studied two CME events at comet CG in September 2014 and noticed 
a jump in the magnetic field strength as well as ion energy and increased ion
count rates associated with the two CME events. 
\cite{Edberg_2016JGR} investigated the interactions between
four Corotating Interaction Regions (CIRs) and the comet and found
compressions of the plasma in the inner coma, which can increase fluxes
of suprathermal electrons and hence electron-impact ionization rates of the neutrals and elevated plasma
densities, as well as the magnetic field strengths.
\cite{Edberg_2016MNRAS} studied a CME impact on comet CG in September 2015 and
found a huge compression, which
increases the flux of suprathermal electrons, the plasma density, and
the magnetic field strength. They also reported
unprecedentedly large magnetic field spikes at a distance of 800\,km
away from the nucleus and interpreted them as magnetic flux ropes,
which are possibly formed by magnetic reconnection in the coma or
strong shears causing Kelvin-Helmholtz instabilities in the plasma flow.
\cite{Goetz_2019} reported that a significant increase in the magnetic field magnitude
to about 300 nT was observed on July 3, 2015, which is the strongest magnetic field 
ever measured in the plasma environment of a comet. They used ENLIL \citep{Odstrcil_2003} to 
simulate the solar wind conditions during that period at comet CG and 
concluded that the unusual behavior was caused 
by the impact of a combination of an interplanetary CME and a CIR.

Numerical modeling has been frequently applied to study the cometary magnetosphere.
There are three major approaches: the fluid approximation \citep{Gombosi_1996, Hansen_2007, Rubin_2014,
Rubin_2015, Huang_2016, Baranov_2019}, the hybrid models \citep{Bagdonat_2002,
Koenders_2015, Wedlund_2017, Lindkvist_2018, Alho_2019}, and fully kinetic models \citep{Deca_2017, Deca_2019, 
Gunell_2019, Divin_2020,Beth_2022}. 
In the fluid approach, the plasma is
treated as one or multiple fluids, and their motion is described by the
magnetohydrodynamic (MHD) equations, while in the hybrid approach, the
ions are treated as individual particles, and the electrons are
treated as a massless fluid. 
Recently, fully kinetic (sometimes called Particle-In-Cell, PIC) models, 
which treat both ions and electrons as particles, have been developed \citep{Deca_2017, Deca_2019, 
Gunell_2019, Divin_2020} to study the kinetic behaviors of both ions and electrons
in the cometary magnetosphere. All these models have successfully modeled
the cometary plasma environment, and the simulation results generally agree well 
with observations. However, there are limited attempts to simulate the impact of space weather events on a comet,
mainly due to the high computational cost required for running long-duration simulations to cover the passage of a CME over the cometary magnetosphere. \cite{Jia_2007} simulated the crossing process of a typical heliospheric current sheet at a comet with their single fluid MHD model and showed that disconnection events can form in the cometary plasma tail. In a similar approach, \cite{Jia_2009} simulated the interaction between a CME and a comet and showed
that the CME can trigger tail disconnection events. 
In this manuscript, we use
our state-of-the-art 3-D multi-fluid plasma-neutral interaction model to simulate
the interaction between a CME and comet CG and, investigate the cometary magnetospheric response, especially in the inner coma environment, and compare the results with Rosetta observations.

\section{Methodology}

The  3-D multi-fluid plasma-neutral interaction model has been used to simulate the
global plasma boundaries of comet CG \citep{Huang_2016} as well as understanding
the diamagnetic field regions \citep{Huang_2016MNRAS, Huang_2018}. 
\cite{Rubin_2014} have shown that multi-fluid plasma models can capture the effect of gyration of different ion fluids,
and the simulation results, within the limitations of the fluid approach, show good agreement with a hybrid model \citep{Muller_2011}.
In addition, our multi-fluid MHD model has been applied in simulating the  plasma environment
of Mars \citep{Najib_2011, Bougher_2015, Ma_2011,Dong_2018,Dong_2018b} and Europa \citep{Rubin_2015, Jia_2018, Harris_2021}. These studies demonstrated that the model accurately describes the plasma environment, as confirmed by comparisons with in-situ observations.

Here, we briefly describe our state-of-the-art 3-D multi-fluid
plasma-neutral interaction model. A more detailed discussion can be found
in \cite{Huang_2016, Huang_2016MNRAS, Huang_2018}. Four different
fluids are simulated in the model: one for the cometary neutral gas,
two for ions (cometary water ions and solar wind protons), and the last one for electrons. 
The cometary neutral gas fluid is described by the Euler equations as it does not
interact with the magnetic and electric fields. In contrast, the plasma (ions and electrons) are represented by the multi-ion MHD equations
\citep{Rubin_2014,Huang_2016}. The source and loss of the ions and electrons are adequately described by elastic and inelastic collisions between different fluid particles, 
various chemical reactions (e.g., photo-ionization, electron impact ionization of the cometary neutral gas),
charge exchange between neutrals and ions, as well as ion-electron recombination. The
magnetic field is obtained from Faraday's law, neglecting the 
Hall velocity and the electron pressure gradient term in the induction equation. 

The Euler and multi-ion MHD equations can describe the behavior and interactions
of the cometary neutral gas, the cometary ions, the solar wind protons, and the electrons self-consistently. 
These equations are solved by the BATS-R-US (Block-Adaptive Tree Solarwind Roe-type Upwind Scheme) code
\citep{Powell_1999, Toth_2012} within the Space Weather Modeling Framework 
(SWMF, \cite{Toth_2005, Toth_2012, Gombosi_2021}) on a 3D block adaptive grid. 
The adaptive grid can resolve different length scales in a global system, which is critical
for studying the interactions between solar wind and comet CG. The nucleus in the model
can be either an idealized comet with a spherical shape or the realistic shape model of comet CG \citep{Preusker_2015}. 
At the inner boundary, the neutral gas outflow is illumination-driven based on the empirical relation suggested by
\cite{Davidsson_2007, Tenishev_2008}. The ions, electrons, and magnetic field
follow a reflected boundary condition, meaning they cannot penetrate the body. 
At the outer boundaries, a floating boundary condition (zero gradient) is applied, 
except for the upstream boundary, where the solar wind plasma and the interplanetary magnetic field (IMF) conditions are prescribed.

This work focuses on the situation where comet CG is near its perihelion at 1.3\,AU. We use the same photo-ionization rate (6$\times$10$^{-7}$\,s$^{-1}$) as in \cite{Huang_2016}. We apply a spherical body with the radius of 2\,km, and the neutral gas outflow is illumination-driven to take into account the day-night asymmetry.
The computational domain extends $\pm1.05\times10^6\,$km in the $X$ direction,
and $\pm5.24\times10^5$\,km in both $Y$ and $Z$ directions, with
18 levels of refinements, increasing the resolution by a factor of two at every level. 
The smallest cell is 0.125\,km near the nucleus, 
while the largest cell is about 16,384\,km far away from the comet. 
There are about 26,000 grid blocks with 1.64 million grid cells. We use the
body-centered Solar EQuatorial (CSEQ) coordinate system, so the upstream boundary condition is specified in the $+X$ direction.

\begin{figure}[ht!]
\center
\includegraphics[width=0.5\linewidth]{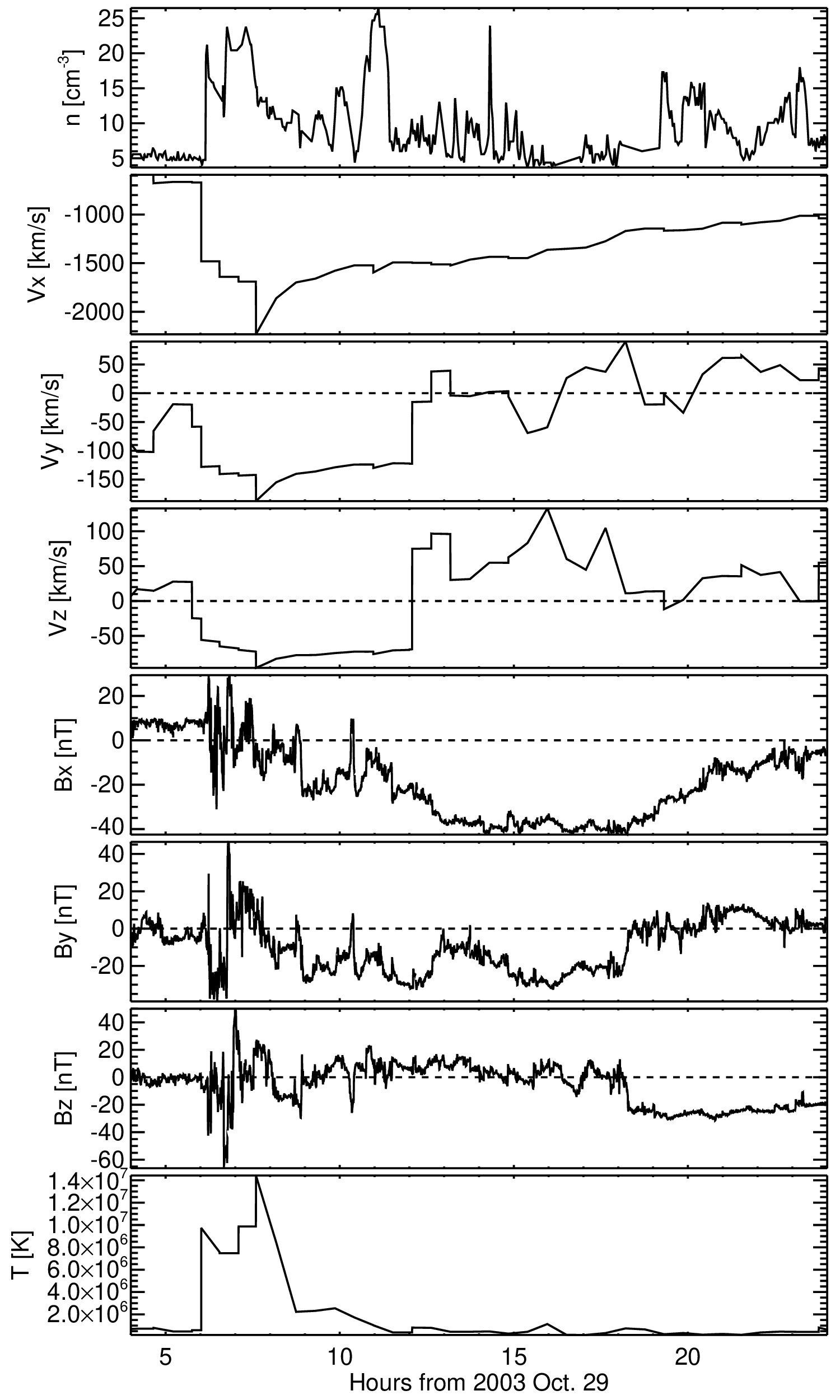}
\caption{This figure shows the solar wind and IMF conditions for the Halloween event.}
\label{fig:sw}
\end{figure}

In order to simulate the interaction between a CME and comet CG, the upstream solar wind and IMF conditions need to be prescribed as input. 
Unfortunately, RPC did not provide solar wind measurements because 
the Rosetta spacecraft remained within the cometary magnetosphere while in orbit around the comet and was typically pointed in a very different direction than the Sun. One of the motivations of the study is to investigate  if an extreme CME can produce the unusually high magnetic 
fields (about 300\,nT) in the inner coma reported in \cite{Goetz_2019}, so
we use the solar wind observation during the 2003 Halloween event, which is 
one of the strongest CMEs observed in recent decades and caused one of the strongest 
geomagnetic storms \citep{Pulkkinen_2013}.  To save computational cost, we only simulate the interaction of the comet with the solar wind and IMF measured at Earth
between 2003 October 29, 04:00 UT and October 30, 00:00 UT, corresponding to a total duration of 20 hours. The solar wind data is directly obtained from CCMC, which is reconstructed from 
several instruments on ACE and Geotail to best represent the solar wind conditions 
due to a large portion of missing observational data from the
ACE solar wind instruments \citep{Pulkkinen_2013}. 
As a crude estimation, B$_y$ and B$_z$ can be scaled as 1/r, while the proton density and Bx can be scaled as 1/r$^2$. For a more accurate description, a model (e.g., MSWIM2D by \cite{Keebler_2022}) is needed to propagate the solar wind from 1 to 1.3\,au. As this study focuses on the cometary magnetospheric response
to an extreme CME, it is not critical to scale the solar wind observations to the exact same heliocentric distance.
Figure~\ref{fig:sw} shows the upstream
solar wind boundary conditions applied in our model.

\section{Simulation results}
\subsection{The Bow Shock}
\begin{figure}[ht!]
\center
\includegraphics[width=0.7\linewidth]{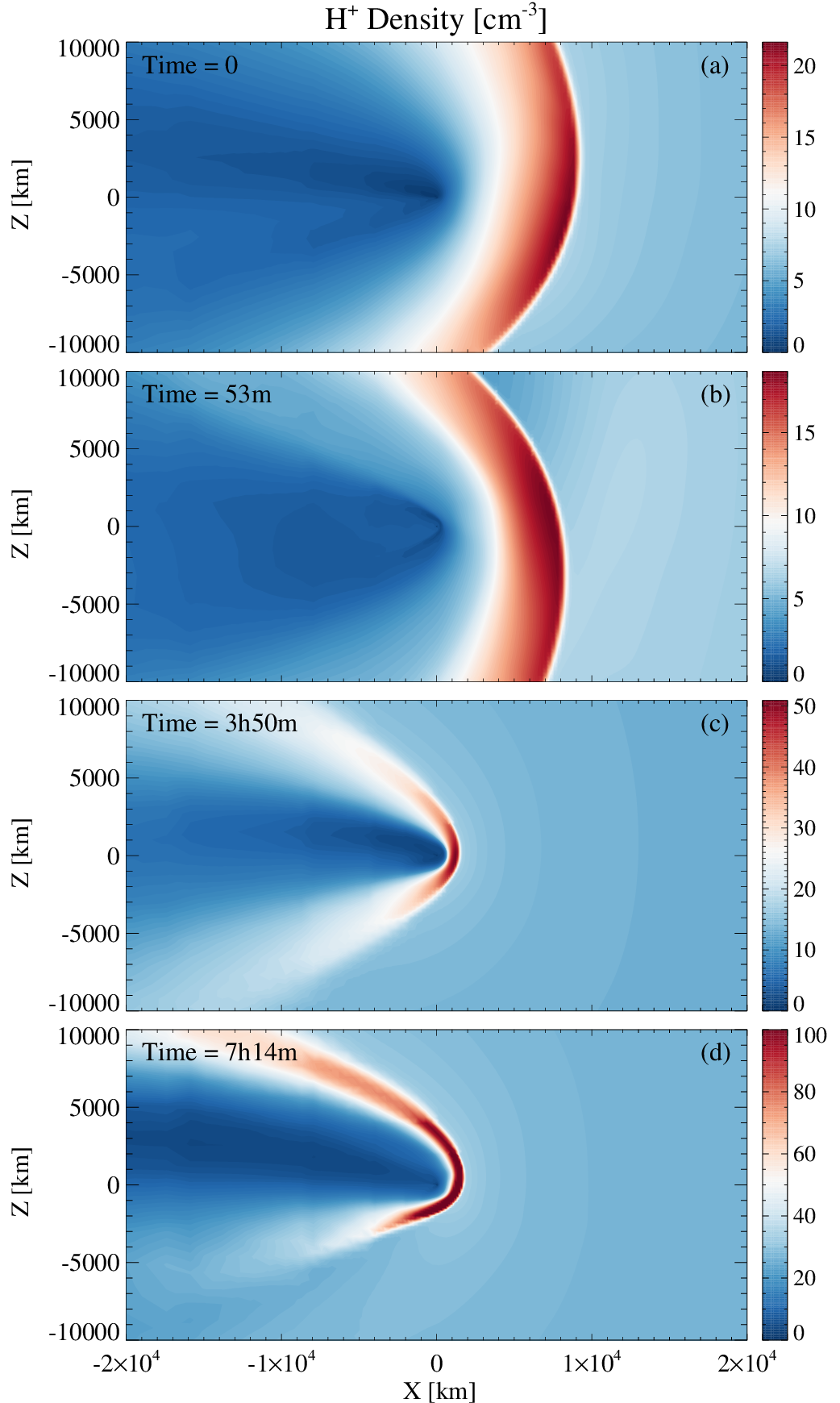}
\caption{This figure shows the simulated H$^+$ density at different simulation times (labeled at the top left corner of each
subplot). The color range is adjusted each time for better visualization of the compression.}
\label{fig:bs}
\end{figure}

Figure~\ref{fig:bs} shows the simulated bow shock at four different
time stamps. It shows that the bow shock changes significantly during the CME event under different solar wind conditions. 
\cite{Huang_2016} suggested that the bow shock is at about 10,000 km for 
a fixed solar wind condition with the illumination-driven neutral gas outflow, which is 
similar to the bow shock location in Panel\,(a) in this simulation. However, the shapes are different: the bow shock in \cite{Huang_2016} is
titled towards -$Z$ direction while the bow shock in Panel\,(a) is titled towards +$Z$ direction. This is caused by the different 
solar wind flows and IMFs: only the $V_\text{x}$ and B$_\text{y}$ component are non-zero ($V_\text{x}$=-400\,km/s and B$_\text{y}$ = 4.8\,nT) in \cite{Huang_2016}
while all three components of the solar wind velocity and IMF are non-zero ($\mathbf{v}$ = [-593, -88, 7.00]\,km/s, $\mathbf{B}$ = [7.65, -5.06, -2.59]\,nT) at the beginning of the current simulation.
In Panel\,(b), the bow shock is tilted slightly towards the -$Z$ direction, similar to the bow
shock in \cite{Huang_2016}. \cite{Koenders_2013} also investigated how the 
bow shock distance changed due to different upstream solar wind parameters
based on a symmetric neutral outflow. Their predicted distances were smaller than the values from our model because we use an asymmetric
neutral outflow driven by solar illumination. Besides, they also discussed the physical mechanics responsible for the discrepancy of bow shock distances obtained from the hybrid model and MHD model under the symmetric neutral background. 
Significant compression can be seen when the CME arrives at the bow shock (Panel\,(c)): 
the bow shock is strongly compressed, moving from 10,000\,km to around 2,000\,km and the
proton number density is much higher ($>$ 100\,cm$^{-3}$) than during quiet time ($<$ 22\,cm$^{-3}$).

\subsection{The Inner Coma}
\label{sec:inner}
\begin{figure}[ht!]
\center
\includegraphics[width=1\linewidth]{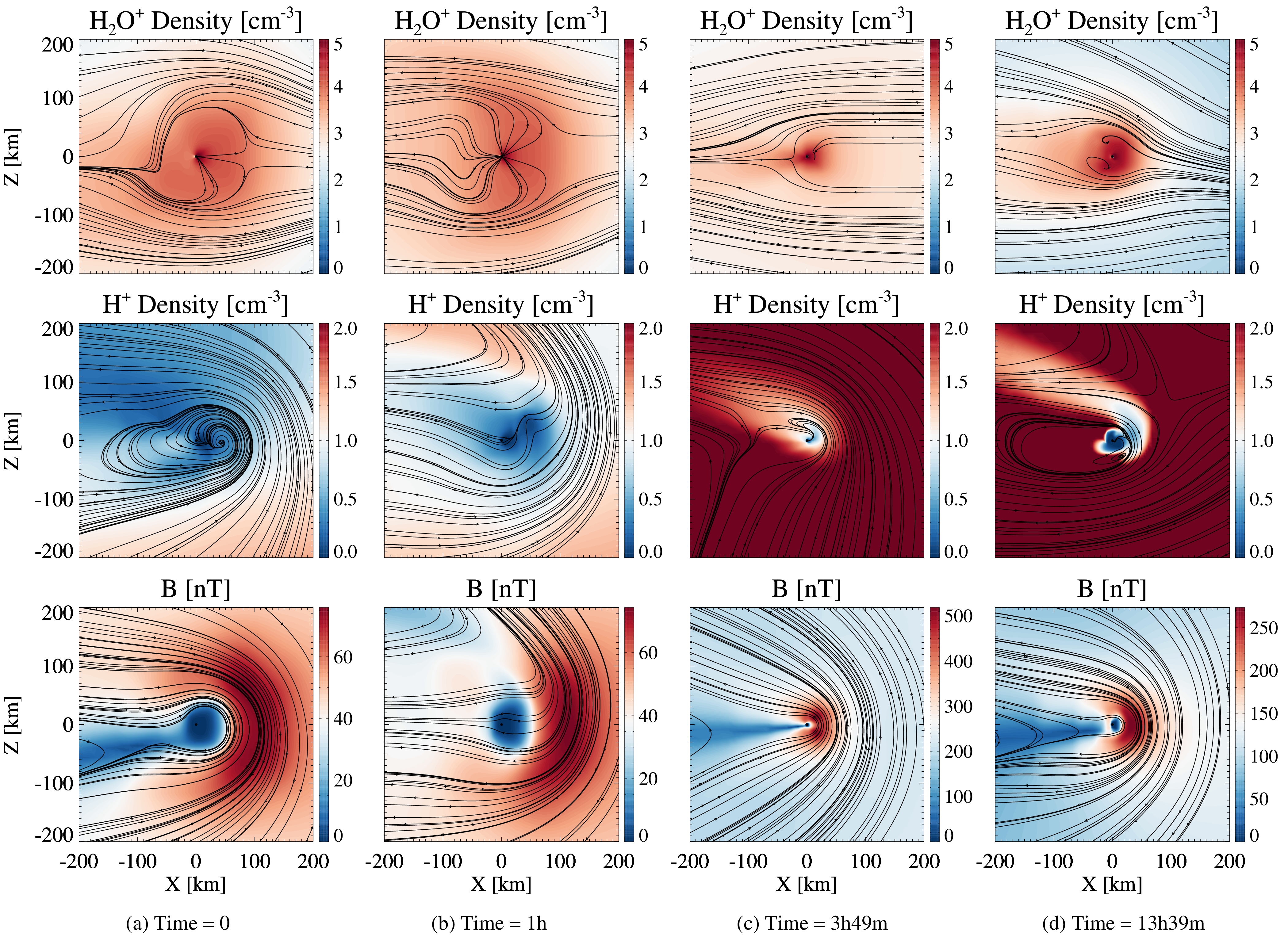}
\caption{This figure shows the simulated H$_2$O$^+$ and H$^+$ densities with the associated velocity streamlines, and the magnetic field magnitude with magnetic field lines in the inner coma in the $Y=0$ plane, within 200\,km of the nucleus
at different simulation times during a Halloween class CME event at comet CG. The simulation time is shown in the 
bottom of each column. The range of the magnetic field is varied for better visualization of the compression.}
\label{fig:yplane}
\end{figure}

\begin{figure}[ht!]
\center
\includegraphics[width=1\linewidth]{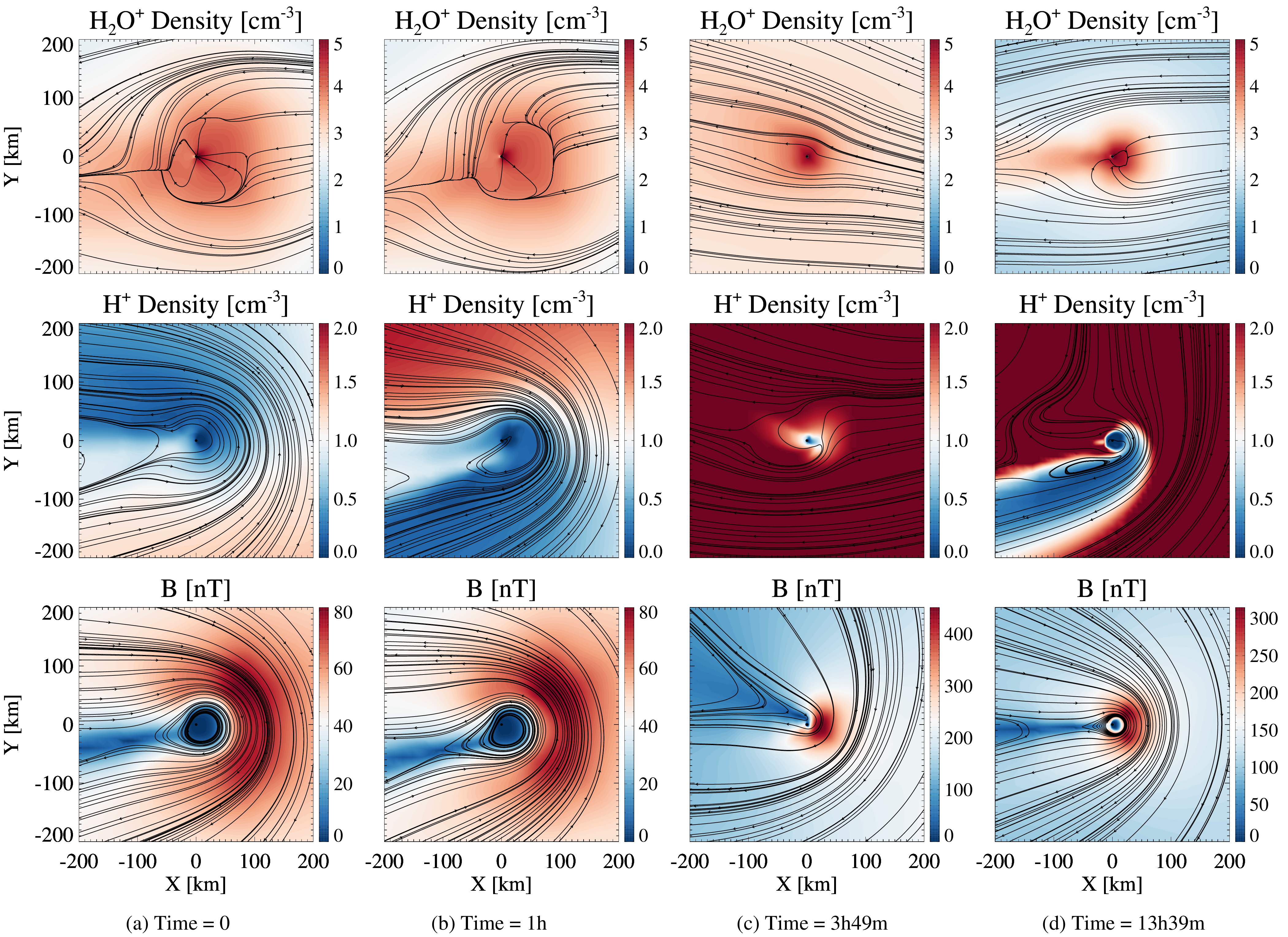}
\caption{This figure shows the same physical quantities as Figure~\ref{fig:yplane} but for the the $Z=0$ plane. }
\label{fig:zplane}
\end{figure}

Figures \ref{fig:yplane} and \ref{fig:zplane} provide the time 
evolution of the impact of the CME on the plasma environment in two perpendicular planes ($Y$=0 and $Z$=0). Both figures show the cometary ion (H$_2$O$^+$) and solar wind proton (H$^+$) densities in the inner coma. Furthermore, the magnetic field magnitude (B) with magnetic field lines are given. We can see that the cometary ion density in the radial distances between 50\,km and 200\,km
in Panels (c) and (d) is smaller than in Panels (a) and (b), but the proton density is higher, which indicates that
the cometary ion density is anti-correlated with the proton density. This anti-correlation
can be explained by the compression of the cometary ion density, which is caused by the higher dynamic pressure of the solar wind. 
This compression is confirmed by the higher cometary ion density within 50\,km in Panels (c) and (d), as indicated by the streamlines of the cometary ions.

The magnetic field is also significantly compressed as the CME
arrives at comet CG: the maximum magnitude just outside the diamagnetic cavity jumps from around 70\,nT
to more than 500\,nT in Panel (c) of Figure~\ref{fig:yplane} and more than
250\,nT in Panel (d) of Figure~\ref{fig:zplane}. 
The diamagnetic cavity almost disappears in Panel (c), which is caused by the extremely
large dynamic pressure of the CME. \cite{Gombosi_2015} provided a rough approximation that the location of the diamagnetic cavity (in the +$X$ direction) is
inversely proportional to the solar wind dynamic pressure
$\rho_{sw} u_{sw}^2$. 
The maximum solar wind speed during the CME event 
is about 2200\,km/s and the corresponding number density is about 13\,cm$^{-3}$, as shown in Figure~\ref{fig:sw}.
Because we use the same parameters for comet CG 
(the heliocentric distance, the total neutral gas outflow, and the collision rates between different fluids) as 
\cite{Huang_2016}, the predicted location of the diamagnetic cavity is about 8 times smaller, 
which is about 12\,km in the +$X$ direction. The size of our simulated diamagnetic cavity is consistent with the theoretical expectation. 

Asymmetries can also be noticed in the inner coma due to different upstream solar wind flows and IMFs. 
In Figure~\ref{fig:yplane}, we can see that the magnetic field lines are slightly draped in the $-Z$ direction in Panel\,(a), while they are draped towards $+Z$ direction in Panel\,(b). 
They are more or less symmetric in Panels\,(c) and (d). In Figure~\ref{fig:zplane}, the magnetic field lines are draped towards the $-Y$ direction in both Panels\,(a) and (b), while they are bent towards $+Y$ direction in Panel\,(c). And they are symmetric in Panel\,(d). 
We can also see in the middle panels of Figure~\ref{fig:zplane} that the solar wind protons are deflected in different directions because of the gyration of the other ion fluids around the charge-averaged ion velocity, as discussed in detail by \cite{Rubin_2014} and \cite{Huang_2016}.

\subsection{Virtual Satellites}

\begin{figure}[ht!]
\center
\includegraphics[width=0.95\linewidth]{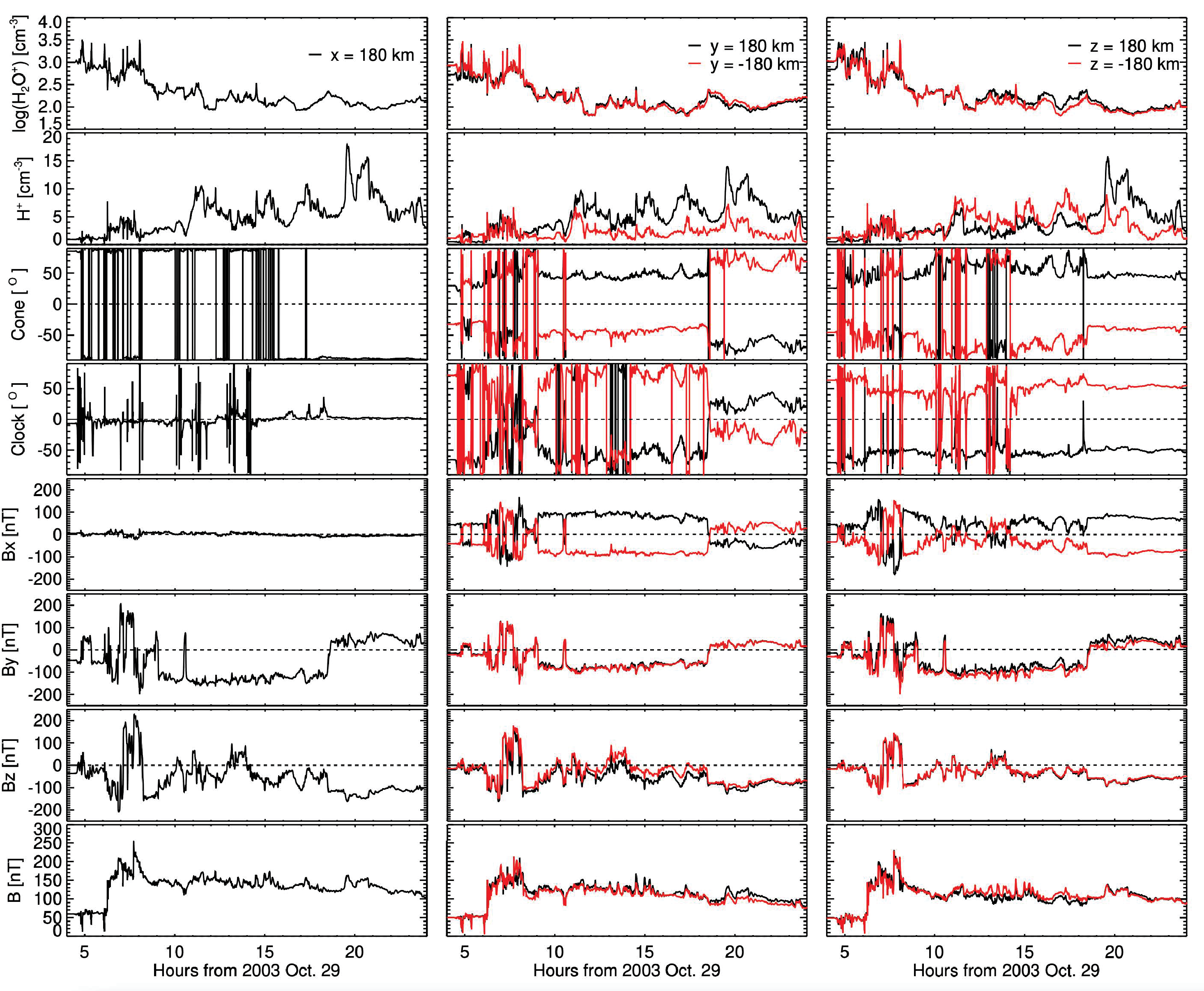}

\caption{This figure shows the simulated H$_2$O$^+$, H$^+$ densities, cone and clock angels, as well as magnetic fields (B$_\text{x}$, B$_\text{y}$, B$_\text{z}$ and the magnitude B) at the virtual satellite locations. The left panels are associated with the virtual satellites at x = 180\,km, while the middle and right panels correspond to the y = $\pm$180\,km and z = $\pm$180\,km, respectively.}
\label{fig:sat}
\end{figure}

We placed five virtual satellites in the inner coma along the three coordinate axes at $x = 180$\,km, $y=\pm 180$\,km and $z=\pm 180$\,km, respectively.
Extracting the model solutions at these virtual satellites
provides predictions of what Rosetta would have measured at those locations during the event. Figure~\ref{fig:sat} shows the simulated 
cometary ion density, proton density, magnetic field cone and clock angles (based on the formulas in \cite{Goetz_2019}), and magnetic fields at the virtual satellite locations. 
Fluctuations of varying amplitude in all quantities can be readily noticed, 
which indicates that there is no stationary state of the plasma environment 
if a dynamic solar wind condition is applied.
The cometary ion and proton densities show similar trends at all virtual satellite locations and 
the anti-correlation 
of solar wind and cometary ion densities
is also found, which is consistent with
the discussion in Section\,\ref{sec:inner}.
Besides, the differences between the solar wind proton densities 
at $y=+180$\,km and $y=-180$\,km (and similar for $z=\pm 180$\,km)
are associated with the deflation
of the solar wind, which is the gyration around the charge-averaged ion velocity, due to different magnetic fields, as indicated by the cone and clock angles. 
We also notice a cometary ion density increase at around 
8 hours of simulation time, which may be associated with the 
increased electron impact ionization due to the high
solar wind temperature, as suggested in \cite{Edberg_2016MNRAS,Edberg_2016JGR}.

The B$_\text{y}$ and B$_\text{z}$ components show similar trends at those locations during the event, while
the B$_\text{x}$ component varies among different locations.  
The magnetic field lines are bent to surround the diamagnetic cavity, so B$_\text{x}$ should be 0 along the x-axis 
in the upstream direction of the diamagnetic cavity, which explains why B$_\text{x}$ is almost 0 at $x=180$\,km. The distinct time evolution of the B$_\text{x}$ component at $y=+180$\,km and $y=-180$\,km (and similar for $z=\pm 180$\,km) is caused by the different upstream
IMF and how those field lines are bent around the diamagnetic cavity, as discussed in Section\,\ref{sec:inner}. For the same reason, 
B$_\text{x}$ is asymmetric around $y=\pm 180$\,km
and $z=\pm 180$\,km, as shown in Figure~\ref{fig:sat}.
The reversal of B$_\text{x}$ and B$_\text{y}$ is also noticed by \cite{Goetz_2019}; however, B$_\text{z}$ shows different behavior, which
may come from the different magnetic field structures, as they suggested their 
observations were caused by a CME combined with a CIR.
The maximum magnitude of the magnetic field is
around 250\,nT, which is slightly less than what \cite{Goetz_2019} observed.

\section{Summary and Discussions}

This paper simulates the interaction between a Halloween-class CME and comet CG. We apply the
solar wind data for the Halloween event between 2003 October 29 04:00 UT and October 30 00:00 UT
from CCMC and use the 3-D multi-fluid plasma-neutral gas interaction model \citep{Huang_2016}. Comet CG is treated as
an idealized comet with a spherical shape, and the neutral gas outflow is driven by solar illumination. Our results
show that the cometary bow shock can be significantly compressed, and its standoff distance is reduced from around 10,000\,km
to about 2,000\,km when the Halloween class CME hits comet CG. We also discover that the cometary ion density in the region between
50\,km and 200\,km is anti-correlated with the solar wind proton density due to the compression caused by the solar wind.

The maximum magnitude of the magnetic field as observed at three virtual satellites at $x/y/z=180$\,km is about 250\,nT, which 
is slightly less than the 300\,nT observation reported by \cite{Goetz_2019}. The Halloween class CME is one of the strongest CMEs
observed in the last few decades, with propagation speed exceeding 2,200\,km/s. The standoff distance of the diamagnetic cavity is inversely proportional to the solar wind dynamic pressure (${n_{sw} u_{sw}^2}$),
as suggested by \cite{Gombosi_2015}. 
A CME with a similar magnetic field strength but 
smaller dynamic pressure will cause less compression in the inner coma, leading to a larger standoff distance of the diamagnetic
cavity. In such a case, the interaction may produce the 300\,nT value observed by \cite{Goetz_2019}
because the most significant magnetic field magnitude is more than 500\,nT just outside the diamagnetic cavity in our simulation (See Panel\,(c) in Figure~\ref{fig:yplane}).

Nearly all previous numerical simulations used a steady solar wind flow \citep{Hansen_2007, Rubin_2014,
Rubin_2015, Huang_2016,Wedlund_2017, Lindkvist_2018, Alho_2019}. 
\cite{Jia_2007} applied approximated current sheet configurations, while \cite{Jia_2009} used a rotation of the IMF vector as a simplified flux rope. 
This is the first time realistic solar wind data has been used to simulate the solar wind-comet interaction.
Our results show ubiquitous fluctuations in the cometary ion density, proton density, and magnetic field, which suggest
that the plasma environment is not steady. Furthermore, our simulation assumed an idealized comet with a spherical body and illumination-driven neutral gas outflow to save computational costs. A more realistic and dynamic plasma environment is expected if the real shape (e.g., SHAP5.1 \citep{Preusker_2015}, which was previously employed in \cite{Huang_2016}), and the comet's rotation were included in the simulation.

We find that the plasma structures are asymmetric during the event due to the time-varying upstream solar wind flow and IMF. \cite{Huang_2018} showed that plasma structures are highly asymmetric due to the Hall current along the diamagnetic cavity boundary. They showed that the Hall current can cause magnetic reconnection on the day-side, which generates weak magnetic field regions outside the global diamagnetic cavity. We expect that the Hall effect would further enhance the asymmetry and cause more complex plasma structures in the inner coma if it is taken into account. However, a Hall MHD simulation is more than 50 times more expensive than a simulation without the Hall effect. With limited resources after the Rosetta mission was concluded, it is impossible to carry out such a costly simulation for the CME interaction with comet CG. However, we plan to carry out a similar simulation with the Hall effect included to investigate how it will change the inner coma environment for different IMF conditions in the future.

\begin{acknowledgments}
The authors would like to thank the ROSINA team for supporting this research. 
The authors also thank the ESA Rosetta team for providing the opportunities to
study this unique comet and their continuous support. 

We acknowledge the high-performance computing support from Cheyenne
(doi:10.5065/D6RX99HX) provided by NCAR's Computational and Information Systems Laboratory,
sponsored by the NSF, and the computation time on Frontera (doi:10.1145/3311790.3396656)
sponsored by NSF and the NASA supercomputing system Pleiades. 

Work by M.R. at the University of Bern was funded by the Canton of Bern and the Swiss National Science Foundation (SNSF; 200020\_207312).

\end{acknowledgments}

\newpage
\bibliography{reference}

\begin{thebibliography}{58}
\expandafter\ifx\csname natexlab\endcsname\relax\def\natexlab#1{#1}\fi

\bibitem[{{Alho} {et~al.}(2019){Alho}, {Simon Wedlund}, {Nilsson}, {Kallio},
  {Jarvinen}, \& {Pulkkinen}}]{Alho_2019}
{Alho}, M., {Simon Wedlund}, C., {Nilsson}, H., {et~al.} 2019, \aap, 630, A45

\bibitem[{{Bagdonat} \& {Motschmann}(2002)}]{Bagdonat_2002}
{Bagdonat}, T., \& {Motschmann}, U. 2002, Earth Moon and Planets, 90, 305

\bibitem[{{Baranov} {et~al.}(2019){Baranov}, {Alexashov}, \&
  {Lebedev}}]{Baranov_2019}
{Baranov}, V.~B., {Alexashov}, D.~B., \& {Lebedev}, M.~G. 2019, \mnras, 482,
  5642

\bibitem[{{Beth} {et~al.}(2022){Beth}, {Gunell}, {Simon Wedlund}, {Goetz},
  {Nilsson}, \& {Hamrin}}]{Beth_2022}
{Beth}, A., {Gunell}, H., {Simon Wedlund}, C., {et~al.} 2022, \aap, 667, A143

\bibitem[{{Bougher} {et~al.}(2015){Bougher}, {Jakosky}, {Halekas}, {Grebowsky},
  {Luhmann}, {Mahaffy}, {Connerney}, {Eparvier}, {Ergun}, {Larson}, {McFadden},
  {Mitchell}, {Schneider}, {Zurek}, {Mazelle}, {Andersson}, {Andrews}, {Baird},
  {Baker}, {Bell}, {Benna}, {Brain}, {Chaffin}, {Chamberlin}, {Chaufray},
  {Clarke}, {Collinson}, {Combi}, {Crary}, {Cravens}, {Crismani}, {Curry},
  {Curtis}, {Deighan}, {Delory}, {Dewey}, {DiBraccio}, {Dong}, {Dong}, {Dunn},
  {Elrod}, {England}, {Eriksson}, {Espley}, {Evans}, {Fang}, {Fillingim},
  {Fortier}, {Fowler}, {Fox}, {Gr{\"o}ller}, {Guzewich}, {Hara}, {Harada},
  {Holsclaw}, {Jain}, {Jolitz}, {Leblanc}, {Lee}, {Lee}, {Lefevre}, {Lillis},
  {Livi}, {Lo}, {Ma}, {Mayyasi}, {McClintock}, {McEnulty}, {Modolo},
  {Montmessin}, {Morooka}, {Nagy}, {Olsen}, {Peterson}, {Rahmati},
  {Ruhunusiri}, {Russell}, {Sakai}, {Sauvaud}, {Seki}, {Steckiewicz},
  {Stevens}, {Stewart}, {Stiepen}, {Stone}, {Tenishev}, {Thiemann}, {Tolson},
  {Toublanc}, {Vogt}, {Weber}, {Withers}, {Woods}, \& {Yelle}}]{Bougher_2015}
{Bougher}, S., {Jakosky}, B., {Halekas}, J., {et~al.} 2015, Science, 350, 0459

\bibitem[{{Carr} {et~al.}(2007){Carr}, {Cupido}, {Lee}, {Balogh}, {Beek},
  {Burch}, {Dunford}, {Eriksson}, {Gill}, {Glassmeier}, {Goldstein},
  {Lagoutte}, {Lundin}, {Lundin}, {Lybekk}, {Michau}, {Musmann}, {Nilsson},
  {Pollock}, {Richter}, \& {Trotignon}}]{Carr_2007}
{Carr}, C., {Cupido}, E., {Lee}, C.~G.~Y., {et~al.} 2007, \ssr, 128, 629

\bibitem[{{Cravens}(1986)}]{Cravens_1986}
{Cravens}, T.~E. 1986, in ESA Special Publication, Vol. 250, ESLAB Symposium on
  the Exploration of Halley's Comet, ed. B.~{Battrick}, E.~J. {Rolfe}, \&
  R.~{Reinhard}, 241--246

\bibitem[{{Davidsson} {et~al.}(2007){Davidsson}, {Guti{\'e}rrez}, \&
  {Rickman}}]{Davidsson_2007}
{Davidsson}, B.~J.~R., {Guti{\'e}rrez}, P.~J., \& {Rickman}, H. 2007, icarus,
  191, 547

\bibitem[{Deca {et~al.}(2017)Deca, Divin, Henri, Eriksson, Markidis, Olshevsky,
  \& Hor\'anyi}]{Deca_2017}
Deca, J., Divin, A., Henri, P., {et~al.} 2017, Phys. Rev. Lett., 118, 205101

\bibitem[{{Deca} {et~al.}(2019){Deca}, {Henri}, {Divin}, {Eriksson}, {Galand},
  {Beth}, {Ostaszewski}, \& {Hor{\'a}nyi}}]{Deca_2019}
{Deca}, J., {Henri}, P., {Divin}, A., {et~al.} 2019, \prl, 123, 055101

\bibitem[{{Divin} {et~al.}(2020){Divin}, {Deca}, {Eriksson}, {Henri},
  {Lapenta}, {Olshevsky}, \& {Markidis}}]{Divin_2020}
{Divin}, A., {Deca}, J., {Eriksson}, A., {et~al.} 2020, \apjl, 889, L33

\bibitem[{{Dong} {et~al.}(2018{\natexlab{a}}){Dong}, {Lee}, {Ma}, {Lingam},
  {Bougher}, {Luhmann}, {Curry}, {Toth}, {Nagy}, {Tenishev}, {Fang},
  {Mitchell}, {Brain}, \& {Jakosky}}]{Dong_2018b}
{Dong}, C., {Lee}, Y., {Ma}, Y., {et~al.} 2018{\natexlab{a}}, \apjl, 859, L14

\bibitem[{{Dong} {et~al.}(2018{\natexlab{b}}){Dong}, {Bougher}, {Ma}, {Lee},
  {Toth}, {Nagy}, {Fang}, {Luhmann}, {Liemohn}, {Halekas}, {Tenishev},
  {Pawlowski}, \& {Combi}}]{Dong_2018}
{Dong}, C., {Bougher}, S.~W., {Ma}, Y., {et~al.} 2018{\natexlab{b}}, Journal of
  Geophysical Research (Space Physics), 123, 6639

\bibitem[{{Edberg} {et~al.}(2016{\natexlab{a}}){Edberg}, {Alho}, {Andr{\'e}},
  {Andrews}, {Behar}, {Burch}, {Carr}, {Cupido}, {Engelhardt}, {Eriksson},
  {Glassmeier}, {Goetz}, {Goldstein}, {Henri}, {Johansson}, {Koenders},
  {Mandt}, {M{\"o}stl}, {Nilsson}, {Odelstad}, {Richter}, {Wedlund}, {Stenberg
  Wieser}, {Szego}, {Vigren}, \& {Volwerk}}]{Edberg_2016MNRAS}
{Edberg}, N.~J.~T., {Alho}, M., {Andr{\'e}}, M., {et~al.} 2016{\natexlab{a}},
  \mnras, 462, S45

\bibitem[{{Edberg} {et~al.}(2016{\natexlab{b}}){Edberg}, {Eriksson},
  {Odelstad}, {Vigren}, {Andrews}, {Johansson}, {Burch}, {Carr}, {Cupido},
  {Glassmeier}, {Goldstein}, {Halekas}, {Henri}, {Koenders}, {Mandt},
  {Mokashi}, {Nemeth}, {Nilsson}, {Ramstad}, {Richter}, \&
  {Wieser}}]{Edberg_2016JGR}
{Edberg}, N.~J.~T., {Eriksson}, A.~I., {Odelstad}, E., {et~al.}
  2016{\natexlab{b}}, Journal of Geophysical Research (Space Physics), 121, 949

\bibitem[{{Engelhardt} {et~al.}(2018){Engelhardt}, {Eriksson}, {Stenberg
  Wieser}, {Goetz}, {Rubin}, {Henri}, {Nilsson}, {Odelstad}, {Hajra}, \&
  {Valli{\`e}res}}]{Engelhardt_2018}
{Engelhardt}, I.~A.~D., {Eriksson}, A.~I., {Stenberg Wieser}, G., {et~al.}
  2018, \mnras, 477, 1296

\bibitem[{{Galeev} {et~al.}(1985){Galeev}, {Cravens}, \&
  {Gombosi}}]{Galeev_1985}
{Galeev}, A.~A., {Cravens}, T.~E., \& {Gombosi}, T.~I. 1985, \apj, 289, 807

\bibitem[{{Glassmeier} {et~al.}(2007){Glassmeier}, {Boehnhardt}, {Koschny},
  {K{\"u}hrt}, \& {Richter}}]{Glassmeier_2007_Rosetta}
{Glassmeier}, K.-H., {Boehnhardt}, H., {Koschny}, D., {K{\"u}hrt}, E., \&
  {Richter}, I. 2007, Space Sci. Rev., 128, 1

\bibitem[{{Goetz} {et~al.}(2016{\natexlab{a}}){Goetz}, {Koenders}, {Richter},
  {Altwegg}, {Burch}, {Carr}, {Cupido}, {Eriksson}, {G{\"u}ttler}, {Henri},
  {Mokashi}, {Nemeth}, {Nilsson}, {Rubin}, {Sierks}, {Tsurutani}, {Vallat},
  {Volwerk}, \& {Glassmeier}}]{Goetz_2016}
{Goetz}, C., {Koenders}, C., {Richter}, I., {et~al.} 2016{\natexlab{a}}, \aap,
  588, A24

\bibitem[{{Goetz} {et~al.}(2016{\natexlab{b}}){Goetz}, {Koenders}, {Hansen},
  {Burch}, {Carr}, {Eriksson}, {Fr{\"u}hauff}, {G{\"u}ttler}, {Henri},
  {Nilsson}, {Richter}, {Rubin}, {Sierks}, {Tsurutani}, {Volwerk}, \&
  {Glassmeier}}]{Goetz_2016_2}
{Goetz}, C., {Koenders}, C., {Hansen}, K.~C., {et~al.} 2016{\natexlab{b}},
  \mnras, 462, S459

\bibitem[{{Goetz} {et~al.}(2019){Goetz}, {Tsurutani}, {Henri}, {Volwerk},
  {Behar}, {Edberg}, {Eriksson}, {Goldstein}, {Mokashi}, {Nilsson}, {Richter},
  {Wellbrock}, \& {Glassmeier}}]{Goetz_2019}
{Goetz}, C., {Tsurutani}, B.~T., {Henri}, P., {et~al.} 2019, \aap, 630, A38

\bibitem[{{Goetz} {et~al.}(2022){Goetz}, {Behar}, {Beth}, {Bodewits},
  {Bromley}, {Burch}, {Deca}, {Divin}, {Eriksson}, {Feldman}, {Galand},
  {Gunell}, {Henri}, {Heritier}, {Jones}, {Mandt}, {Nilsson}, {Noonan},
  {Odelstad}, {Parker}, {Rubin}, {Simon Wedlund}, {Stephenson}, {Taylor},
  {Vigren}, {Vines}, \& {Volwerk}}]{Goetz_2022}
{Goetz}, C., {Behar}, E., {Beth}, A., {et~al.} 2022, \ssr, 218, 65

\bibitem[{Gombosi(2015)}]{Gombosi_2015}
Gombosi, T.~I. 2015, Physics of Cometary Magnetospheres (John Wiley \& Sons,
  Inc), 169--188

\bibitem[{{Gombosi} {et~al.}(1996){Gombosi}, {De Zeeuw}, {H{\"a}berli}, \&
  {Powell}}]{Gombosi_1996}
{Gombosi}, T.~I., {De Zeeuw}, D.~L., {H{\"a}berli}, R.~M., \& {Powell}, K.~G.
  1996, \jgr, 101, 15233

\bibitem[{{Gombosi} {et~al.}(2021){Gombosi}, {Chen}, {Glocer}, {Huang}, {Jia},
  {Liemohn}, {Manchester}, {Pulkkinen}, {Sachdeva}, {Al Shidi}, {Sokolov},
  {Szente}, {Tenishev}, {Toth}, {van der Holst}, {Welling}, {Zhao}, \&
  {Zou}}]{Gombosi_2021}
{Gombosi}, T.~I., {Chen}, Y., {Glocer}, A., {et~al.} 2021, Journal of Space
  Weather and Space Climate, 11, 42

\bibitem[{{Gunell} {et~al.}(2019){Gunell}, {Lindkvist}, {Goetz}, {Nilsson}, \&
  {Hamrin}}]{Gunell_2019}
{Gunell}, H., {Lindkvist}, J., {Goetz}, C., {Nilsson}, H., \& {Hamrin}, M.
  2019, \aap, 631, A174

\bibitem[{{Gunell} {et~al.}(2018){Gunell}, {Goetz}, {Simon Wedlund},
  {Lindkvist}, {Hamrin}, {Nilsson}, {Llera}, {Eriksson}, \&
  {Holmstr{\"o}m}}]{Gunell_2018}
{Gunell}, H., {Goetz}, C., {Simon Wedlund}, C., {et~al.} 2018, \aap, 619, L2

\bibitem[{{Hansen} {et~al.}(2007){Hansen}, {Bagdonat}, {Motschmann},
  {Alexander}, {Combi}, {Cravens}, {Gombosi}, {Jia}, \&
  {Robertson}}]{Hansen_2007}
{Hansen}, K.~C., {Bagdonat}, T., {Motschmann}, U., {et~al.} 2007, Space Sci.
  Rev., 128, 133

\bibitem[{{Harris} {et~al.}(2021){Harris}, {Jia}, {Slavin}, {Toth}, {Huang}, \&
  {Rubin}}]{Harris_2021}
{Harris}, C. D.~K., {Jia}, X., {Slavin}, J.~A., {et~al.} 2021, Journal of
  Geophysical Research (Space Physics), 126, e28888

\bibitem[{{Henri} {et~al.}(2017){Henri}, {Valli{\`e}res}, {Hajra}, {Goetz},
  {Richter}, {Glassmeier}, {Galand}, {Rubin}, {Eriksson}, {Nemeth}, {Vigren},
  {Beth}, {Burch}, {Carr}, {Nilsson}, {Tsurutani}, \& {Wattieaux}}]{Henri_2017}
{Henri}, P., {Valli{\`e}res}, X., {Hajra}, R., {et~al.} 2017, \mnras, 469, S372

\bibitem[{{Huang} {et~al.}(2016{\natexlab{a}}){Huang}, {T{\'o}th}, {Gombosi},
  {Bieler}, {Combi}, {Hansen}, {Jia}, {Fougere}, {Shou}, {Cravens}, {Tenishev},
  {Altwegg}, \& {Rubin}}]{Huang_2016MNRAS}
{Huang}, Z., {T{\'o}th}, G., {Gombosi}, T.~I., {et~al.} 2016{\natexlab{a}},
  \mnras, 462, S468

\bibitem[{{Huang} {et~al.}(2016{\natexlab{b}}){Huang}, {T{\'o}th}, {Gombosi},
  {Jia}, {Rubin}, {Fougere}, {Tenishev}, {Combi}, {Bieler}, {Hansen}, {Shou},
  \& {Altwegg}}]{Huang_2016}
---. 2016{\natexlab{b}}, Journal of Geophysical Research (Space Physics), 121,
  4247

\bibitem[{{Huang} {et~al.}(2018){Huang}, {T{\'o}th}, {Gombosi}, {Jia}, {Combi},
  {Hansen}, {Fougere}, {Shou}, {Tenishev}, {Altwegg}, \& {Rubin}}]{Huang_2018}
---. 2018, \mnras, 475, 2835

\bibitem[{{Jia} {et~al.}(2018){Jia}, {Kivelson}, {Khurana}, \&
  {Kurth}}]{Jia_2018}
{Jia}, X., {Kivelson}, M.~G., {Khurana}, K.~K., \& {Kurth}, W.~S. 2018, Nature
  Astronomy, 2, 459

\bibitem[{{Jia} {et~al.}(2007){Jia}, {Combi}, {Hansen}, \&
  {Gombosi}}]{Jia_2007}
{Jia}, Y.-D., {Combi}, M.~R., {Hansen}, K.~C., \& {Gombosi}, T.~I. 2007,
  Journal of Geophysical Research (Space Physics), 112, A05223

\bibitem[{{Jia} {et~al.}(2009){Jia}, {Russell}, {Jian}, {Manchester}, {Cohen},
  {Vourlidas}, {Hansen}, {Combi}, \& {Gombosi}}]{Jia_2009}
{Jia}, Y.~D., {Russell}, C.~T., {Jian}, L.~K., {et~al.} 2009, \apjl, 696, L56

\bibitem[{{Keebler} {et~al.}(2022){Keebler}, {T{\'o}th}, {Zieger}, \&
  {Opher}}]{Keebler_2022}
{Keebler}, T.~B., {T{\'o}th}, G., {Zieger}, B., \& {Opher}, M. 2022, \apjs,
  260, 43

\bibitem[{{Koenders} {et~al.}(2013){Koenders}, {Glassmeier}, {Richter},
  {Motschmann}, \& {Rubin}}]{Koenders_2013}
{Koenders}, C., {Glassmeier}, K.-H., {Richter}, I., {Motschmann}, U., \&
  {Rubin}, M. 2013, Planet. Space Sci., 87, 85

\bibitem[{{Koenders} {et~al.}(2015){Koenders}, {Glassmeier}, {Richter},
  {Ranocha}, \& {Motschmann}}]{Koenders_2015}
{Koenders}, C., {Glassmeier}, K.-H., {Richter}, I., {Ranocha}, H., \&
  {Motschmann}, U. 2015, Planet. Space Sci., 105, 101

\bibitem[{{Lindkvist} {et~al.}(2018){Lindkvist}, {Hamrin}, {Gunell}, {Nilsson},
  {Simon Wedlund}, {Kallio}, {Mann}, {Pitk{\"a}nen}, \&
  {Karlsson}}]{Lindkvist_2018}
{Lindkvist}, J., {Hamrin}, M., {Gunell}, H., {et~al.} 2018, \aap, 616, A81

\bibitem[{{Ma} {et~al.}(2011){Ma}, {Russell}, {Nagy}, {Toth}, {Dougherty},
  {Wellbrock}, {Coates}, {Garnier}, {Wahlund}, {Cravens}, {Richard}, \&
  {Crary}}]{Ma_2011}
{Ma}, Y.~J., {Russell}, C.~T., {Nagy}, A.~F., {et~al.} 2011, Journal of
  Geophysical Research (Space Physics), 116, A10213

\bibitem[{{Mandt} {et~al.}(2016){Mandt}, {Eriksson}, {Edberg}, {Koenders},
  {Broiles}, {Fuselier}, {Henri}, {Nemeth}, {Alho}, {Biver}, {Beth}, {Burch},
  {Carr}, {Chae}, {Coates}, {Cupido}, {Galand}, {Glassmeier}, {Goetz},
  {Goldstein}, {Hansen}, {Haiducek}, {Kallio}, {Lebreton}, {Luspay-Kuti},
  {Mokashi}, {Nilsson}, {Opitz}, {Richter}, {Samara}, {Szego}, {Tzou},
  {Volwerk}, {Simon Wedlund}, \& {Stenberg Wieser}}]{Mandt_2016}
{Mandt}, K.~E., {Eriksson}, A., {Edberg}, N.~J.~T., {et~al.} 2016, \mnras, 462,
  S9

\bibitem[{{McKenna-Lawlor} {et~al.}(2016){McKenna-Lawlor}, {Ip}, {Jackson},
  {Odstrcil}, {Nieminen}, {Evans}, {Burch}, {Mandt}, {Goldstein}, {Richter}, \&
  {Dryer}}]{McKenna-Lawlor_2016}
{McKenna-Lawlor}, S., {Ip}, W., {Jackson}, B., {et~al.} 2016, Earth Moon and
  Planets, 117, 1

\bibitem[{{M{\"u}ller} {et~al.}(2011){M{\"u}ller}, {Simon}, {Motschmann},
  {Sch{\"u}le}, {Glassmeier}, \& {Pringle}}]{Muller_2011}
{M{\"u}ller}, J., {Simon}, S., {Motschmann}, U., {et~al.} 2011, Computer
  Physics Communications, 182, 946

\bibitem[{{Najib} {et~al.}(2011){Najib}, {Nagy}, {T{\'o}th}, \&
  {Ma}}]{Najib_2011}
{Najib}, D., {Nagy}, A.~F., {T{\'o}th}, G., \& {Ma}, Y. 2011, Journal of
  Geophysical Research (Space Physics), 116, 5204

\bibitem[{{Neubauer} {et~al.}(1986){Neubauer}, {Glassmeier}, {Pohl}, {Raeder},
  {Acuna}, {Burlaga}, {Ness}, {Musmann}, {Mariani}, {Wallis}, {Ungstrup}, \&
  {Schmidt}}]{Neubauer_1986}
{Neubauer}, F.~M., {Glassmeier}, K.~H., {Pohl}, M., {et~al.} 1986, \nat, 321,
  352

\bibitem[{{Noonan} {et~al.}(2018){Noonan}, {Stern}, {Feldman}, {Broiles},
  {Wedlund}, {Edberg}, {Schindhelm}, {Parker}, {Keeney}, {Vervack}, {Steffl},
  {Knight}, {Weaver}, {Feaga}, {A'Hearn}, \& {Bertaux}}]{Noonan_2018}
{Noonan}, J.~W., {Stern}, S.~A., {Feldman}, P.~D., {et~al.} 2018, \aj, 156, 16

\bibitem[{{Odelstad} {et~al.}(2018){Odelstad}, {Eriksson}, {Johansson},
  {Vigren}, {Henri}, {Gilet}, {Heritier}, {Valli{\`e}res}, {Rubin}, \&
  {Andr{\'e}}}]{Odelstad_2018}
{Odelstad}, E., {Eriksson}, A.~I., {Johansson}, F.~L., {et~al.} 2018, Journal
  of Geophysical Research (Space Physics), 123, 5870

\bibitem[{{Odstrcil}(2003)}]{Odstrcil_2003}
{Odstrcil}, D. 2003, Advances in Space Research, 32, 497

\bibitem[{{Powell} {et~al.}(1999){Powell}, {Roe}, {Linde}, {Gombosi}, \& {de
  Zeeuw}}]{Powell_1999}
{Powell}, K.~G., {Roe}, P.~L., {Linde}, T.~J., {Gombosi}, T.~I., \& {de Zeeuw},
  D.~L. 1999, Journal of Computational Physics, 154, 284

\bibitem[{{Preusker} {et~al.}(2015){Preusker}, {Scholten}, {Matz}, {Roatsch},
  {Willner}, {Hviid}, {Knollenberg}, {Jorda}, {Guti{\'e}rrez}, {K{\"u}hrt},
  {Mottola}, {A'Hearn}, {Thomas}, {Sierks}, {Barbieri}, {Lamy}, {Rodrigo},
  {Koschny}, {Rickman}, {Keller}, {Agarwal}, {Barucci}, {Bertaux}, {Bertini},
  {Cremonese}, {Da Deppo}, {Davidsson}, {Debei}, {De Cecco}, {Fornasier},
  {Fulle}, {Groussin}, {G{\"u}ttler}, {Ip}, {Kramm}, {K{\"u}ppers}, {Lara},
  {Lazzarin}, {Lopez Moreno}, {Marzari}, {Michalik}, {Naletto}, {Oklay},
  {Tubiana}, \& {Vincent}}]{Preusker_2015}
{Preusker}, F., {Scholten}, F., {Matz}, K.-D., {et~al.} 2015, \aap, 583, A33

\bibitem[{{Pulkkinen} {et~al.}(2013){Pulkkinen}, {Rast{\"a}Tter}, {Kuznetsova},
  {Singer}, {Balch}, {Weimer}, {Toth}, {Ridley}, {Gombosi}, {Wiltberger},
  {Raeder}, \& {Weigel}}]{Pulkkinen_2013}
{Pulkkinen}, A., {Rast{\"a}Tter}, L., {Kuznetsova}, M., {et~al.} 2013, Space
  Weather, 11, 369

\bibitem[{{Rubin} {et~al.}(2014){Rubin}, {Koenders}, {Altwegg}, {Combi},
  {Glassmeier}, {Gombosi}, {Hansen}, {Motschmann}, {Richter}, {Tenishev}, \&
  {T{\'o}th}}]{Rubin_2014}
{Rubin}, M., {Koenders}, C., {Altwegg}, K., {et~al.} 2014, icarus, 242, 38

\bibitem[{{Rubin} {et~al.}(2015){Rubin}, {Jia}, {Altwegg}, {Combi}, {Daldorff},
  {Gombosi}, {Khurana}, {Kivelson}, {Tenishev}, {T{\'o}th}, {Holst}, \&
  {Wurz}}]{Rubin_2015}
{Rubin}, M., {Jia}, X., {Altwegg}, K., {et~al.} 2015, Journal of Geophysical
  Research (Space Physics), 120, 3503

\bibitem[{{Simon Wedlund} {et~al.}(2017){Simon Wedlund}, {Alho}, {Gronoff},
  {Kallio}, {Gunell}, {Nilsson}, {Lindkvist}, {Behar}, {Stenberg Wieser}, \&
  {Miloch}}]{Wedlund_2017}
{Simon Wedlund}, C., {Alho}, M., {Gronoff}, G., {et~al.} 2017, \aap, 604, A73

\bibitem[{{Tenishev} {et~al.}(2008){Tenishev}, {Combi}, \&
  {Davidsson}}]{Tenishev_2008}
{Tenishev}, V., {Combi}, M., \& {Davidsson}, B. 2008, \apj, 685, 659

\bibitem[{{T{\'o}th} {et~al.}(2005){T{\'o}th}, {Sokolov}, {Gombosi}, {Chesney},
  {Clauer}, {de Zeeuw}, {Hansen}, {Kane}, {Manchester}, {Oehmke}, {Powell},
  {Ridley}, {Roussev}, {Stout}, {Volberg}, {Wolf}, {Sazykin}, {Chan}, {Yu}, \&
  {K{\'o}ta}}]{Toth_2005}
{T{\'o}th}, G., {Sokolov}, I.~V., {Gombosi}, T.~I., {et~al.} 2005, Journal of
  Geophysical Research (Space Physics), 110, 12226

\bibitem[{{T{\'o}th} {et~al.}(2012){T{\'o}th}, {van der Holst}, {Sokolov}, {De
  Zeeuw}, {Gombosi}, {Fang}, {Manchester}, {Meng}, {Najib}, {Powell}, {Stout},
  {Glocer}, {Ma}, \& {Opher}}]{Toth_2012}
{T{\'o}th}, G., {van der Holst}, B., {Sokolov}, I.~V., {et~al.} 2012, Journal
  of Computational Physics, 231, 870

\end{thebibliography}

\end{document}